\documentclass[a4paper]{jpconf}
\usepackage{graphicx}
\usepackage{hyperref}
\begin{document}
\title{Bridging the gap between cutting-edge science and school: non-formal education in high-energy physics}

\author{Boris Tom\'a\v{s}ik}

\address{Fakulta jadern\'a a fyzik\'aln\v{e} in\v{z}en\'yrsk\'a, \v{C}esk\'e vysok\'e u\v{c}en\'i technick\'e v Praze, B\v{r}ehov\'a 7, 11519 Praha 1, Czech Republic}
\address{Fakulta pr\'irodn\'ych vied, Univerzita Mateja Bela, Tajovsk\'eho 40, 97401 Bansk\'a Bystrica, Slovakia}
\ead{boris.tomasik@cern.ch}

\author{J\'ulia Kekel\'akov\'a}
\address{Fakulta pr\'irodn\'ych vied, Univerzita Mateja Bela, Tajovsk\'eho 40, 97401 Bansk\'a Bystrica, Slovakia}

\author{Ivan Melo}
\address{Fakulta elektrotechniky a informa\v{c}n\'ych technol\'ogi\'i, \v{Z}ilinsk\'a univerzita, Univerzitn\'a 1, 01026 \v{Z}ilina, Slovakia}

\author{Vojt\v{e}ch Pleskot}
\address{Matematicko-fyzik\'aln\'i fakulta, Univerzita Karlova, V Hole\v{s}ovi\v{c}k\'ach 747/2, 18000 Praha 8, Czech Republic}

\begin{abstract}
We report mainly on the global flagship outreach activity in particle physics: the International Particle Physics Masterclasses. 
It is illustrated on the example of Slovakia and the Czech Republic.
The Masterclasses are described and their long-term impact is studied with the help of a survey among former participants. 
The positive effect of Masterclasses in shifting the  attitude towards science is shown. 
We also discuss the modification of Masterclasses for a pandemic situation. 
Finally, we present CASCADE projects as a way to foster the interest in a field which has been strongly enhanced by previous experience, e.g., at Masterclasses. 
\end{abstract}

\section{Introduction}

There could be permanent debates about  which results of the forefront physics research should become the content knowledge at secondary school level. 
Clearly, one of major questions to be addressed here concerns the applicability of such topics by the secondary-level graduates. 
Nevertheless, certain knowledge of the modern results is useful firstly for its generally interesting and essentially cultural character, and secondly due to the need for educated citizens who make informed democratic decisions. 
Then, a practical issue appears, how to convey the knowledge if the teaching personnel could not yet been trained on the new topics. 
The solution from this situation can be provided by a coordinated campaign of outreach activities performed by the research institutions. High-energy 
physics (HEP), sometimes referred to as particle physics,  is perhaps one of the most visible examples of such activities as it profits from immanent interest from the general public. 

It should be also understood that the time and effort invested into outreach are well recognised and valued within the research community since some time, already. The progress vitally depends on public support and therefore the research community must communicate with the wide society. 
Moreover, there is also need for highly qualified and motivated workforce which must be trained and for which the recruitment is appropriate at secondary-school level. 

In this paper we report on several activities that together contribute to the framework of outreach organised by high-energy community in Slovakia and the Czech Republic. 
We hope that the ideas put forward here may inspire implementations in other active research areas, as well.


\section{International Particle Physics Masterclasses}
\label{s:imc}

The flagship activity are the International Particle Physics Masterclasses (IMC) \cite{Kobel05,Foka13,Bilow14,Cecire17,Bilow22}. 
The main idea is to provide the participating upper-secondary level pupils with the experience of research in high-energy physics. 
The IMC are organised annually each spring since 2005, which was proclaimed the International Year of Physics in commemoration of Einstein's Annus Mirabilis in 1905.
Hence, 2024 will mark the 20th issue of the IMC. 
The events are organised locally at participating research institutions.
Nevertheless, since transnational networking activities are involved, global coordination is necessary, which is ensured by the International Particle Physics Outreach Group (IPPOG)\footnote{\href{https://ippog.org}{https://ippog.org}}. 
This is an international collaboration with representatives from participating countries, major laboratories, and experiments. 
It is hosted by CERN. 

Annually, about 13~000 pupils participate in some of the events that are hosted at more than 200~research institutions worldwide. 

\begin{figure}[t]
\begin{center}
\includegraphics[width=0.8\textwidth]{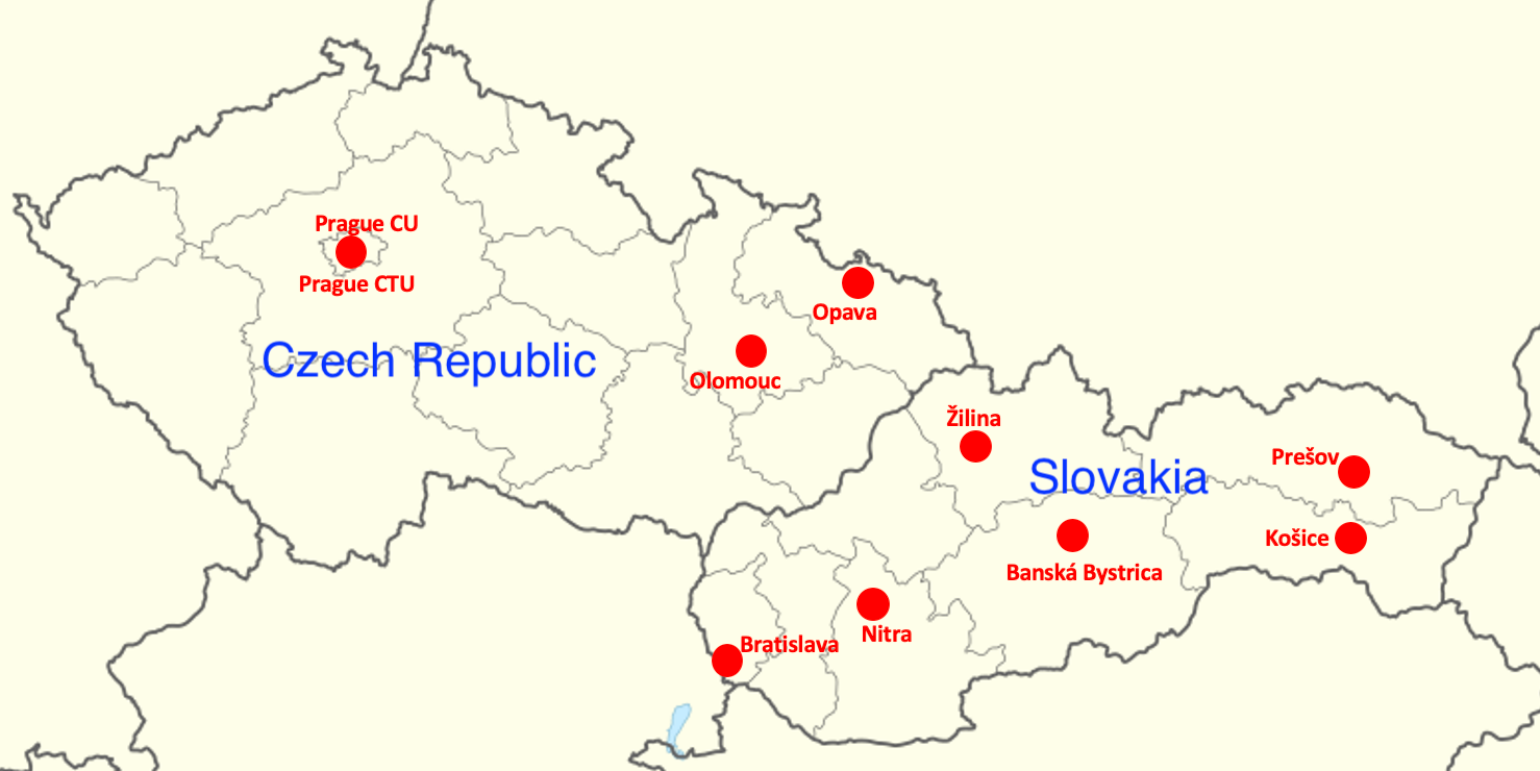}
\end{center}
\caption{\label{f:imcmap}
Places in the Czech Republic and Slovakia, where International Particle Physics Masterclasses have been organised in 2023.}
\end{figure}

In Slovakia and the Czech Republic the geographical coverage with places organising IMC is quite good, 
see Fig.~\ref{f:imcmap}.
%
%
Generally, all big institutions which pursue high-energy research participate. Some smaller places without an active HEP research group also organise the event. In such case, high-energy physicists arrive there from other institution, in order to provide authentic experience. 

The agenda for an IMC event is basically prescribed, with some space to accommodate local needs and uniqueness. 
It is mainly given by the central task which is a hands-on activity to be performed by the participants. 
Since  no prior knowledge relevant to high-energy physics is anticipated, in order to prepare them for this activity, lectures are offered which introduce them to the necessary topics and give them a broad overview of HEP. 
Typically, two different aspects must be addressed: 
Firstly, the theoretical framework known as \emph{Standard Model} of the particle physics, with specific application to the topic of the hands-on exercise. 
Secondly, experimental techniques and data-analysis methods that are  needed for the hands-on exercises. 
In addition to this, there is also space that may be used by the hosting institutions for their presentation, so it can include lab excursions, panel discussions, or any other suitable format. 
The lectures and possibly excursions are usually accommodated in the morning part of the programme. 

This is followed by lunch. Although it sounds like a trivial part of the agenda, it is not! 
On the minimalist side, our experience and evaluation surveys show that poorly organised lunch and/or refreshments can ruin the overall impression from the event and should therefore be given a pertinent attention.  
On the other hand, refreshments provide a great opportunity. If they are organised smartly, e.g., with buffet and finger food, they allow for  easy non- formal discussions between the participants and the scientists. 
Some institutions have good experience with creating such an environment. 
A practical hint: pizzas freshly delivered during a break help to remove conversation barriers and can easily become a highlight of the day, thus generally contributing to very positive experience. 

After the break, there is time for the practical exercise. 
This is prepared by the international collaboration and is provided to any institution that wants to participate. 
Typically, a task of a precisely formulated data analysis is performed. 
Specific software tools are provided to the participants. 
An important feature of the activity is that it is performed on (pre-processed) data that indeed come from some major HEP experiment. 
Currently, all four big experiments working at CERN's Large Hadron Collider (LHC) accelerator have developed and offered exercises based on their data: ALICE, ATLAS, CMS, and LHCb. 
These are all very large experimental collaborations and each of them looks at various aspects of the observed collisions of protons or heavy atomic nuclei. The ATLAS and ALICE collaborations even prepared two exercises each, which address different measurements. 
Other non-CERN experiments have now also worked out exercises based on their own measurements. These include the international BELLE~II collaboration, operating at the Japanese high-energy research organisation KEK, and focusing on rare particles that include b-quark. 
Furthermore, hands-on activity has also been developed by the Pierre Auger collaboration, which run a high-energy cosmic ray detector array in the Argentinian pampas. 
MINER$\nu$A---an experimental collaboration hosted by Fermi National Accelerator Laboratory (Batavia, IL, USA), which  is devoted to studies of neutrinos---has also established their masterclasses exercise. 
Finally, an exercise has been prepared also on the topic of particle therapy, which focuses on setting up irradiation plans for cancer treatment. 
This is an important spin-off which greatly improves the quality of available healthcare. 
Details of all exercises may be found at the dedicated webpage\footnote{\href{https://physicsmasterclasses.org}{https://physicsmasterclasses.org}}.
More exercises can be expected in the future. 

A very important feature of the HEP research is the international collaboration. 
This is illustrated in the agenda by the closing videoconference (VC) that connects up to five participating institutions with moderators who represent one of the major laboratories.
The task of international scheduling of the institutions which connect to the same VC is managed by IPPOG. Often, the VC not only illustrates the international collaboration, but also the need of large statistics in new discoveries. 
In some of the exercises all participants work on the same analysis, but with different data sets. 
Putting together results from all participants during the VC, even from different countries, usually shows an immediate improvement of the obtained signal and  illustrates the need of sufficient statistics and collaboration to achieve it. 

As an example of exercise, let us present  one prepared by the ATLAS collaboration and devoted to the reconstruction of the Z-boson.
This is an unstable neutral particle, which is the mediator of the electro-weak interaction. 
Since it is unstable, it never reaches a detector but decays almost immediately and only (some of) its decay products are measurable. 
Particularly, it may decay into pairs of electron and positron ($e^+e^-$) or pairs of oppositely charged muons ($\mu^+\mu^-$), which can be identified in the detector. 
In such a decay, the (invariant) mass of the decaying parent particle can be determined from the energies $(E_1,E_2)$ and momenta $(\vec p_1,\vec p_2)$ of the decay products as 
\begin{equation}
\label{e:invmass}
M = \frac{1}{c^2} \sqrt{ (E_1+E_2)^2 - (\vec p_1 + \vec p_2)^2c^2}\,  .
\end{equation}
The problem is that electrons and muons can be produced in many other processes and independently from each other.
Applying the formula (\ref{e:invmass}) on such an uncorrelated  pair would lead to a random result. 
However, for pairs that indeed come from the decay of the unstable particle, formula (\ref{e:invmass}) would give the mass of the parent particle.
(There are also other unstable particles than the Z-boson,  that decay into $e^+e^-$ or $\mu^+\mu^-$ pairs.)
Hence, one proceeds by collecting all pairs and filling a histogram of their invariant masses determined from the formula  (\ref{e:invmass}). 
For decaying particles, peaks in the histogram will appear at the values of their masses. 
(In practice, not all pairs are collected, but one applies some acceptance criteria with the aim to enhance the mass peaks over the background due to random pairs.) 
\begin{figure}[t]
\begin{center}
\includegraphics[width=0.65\textwidth]{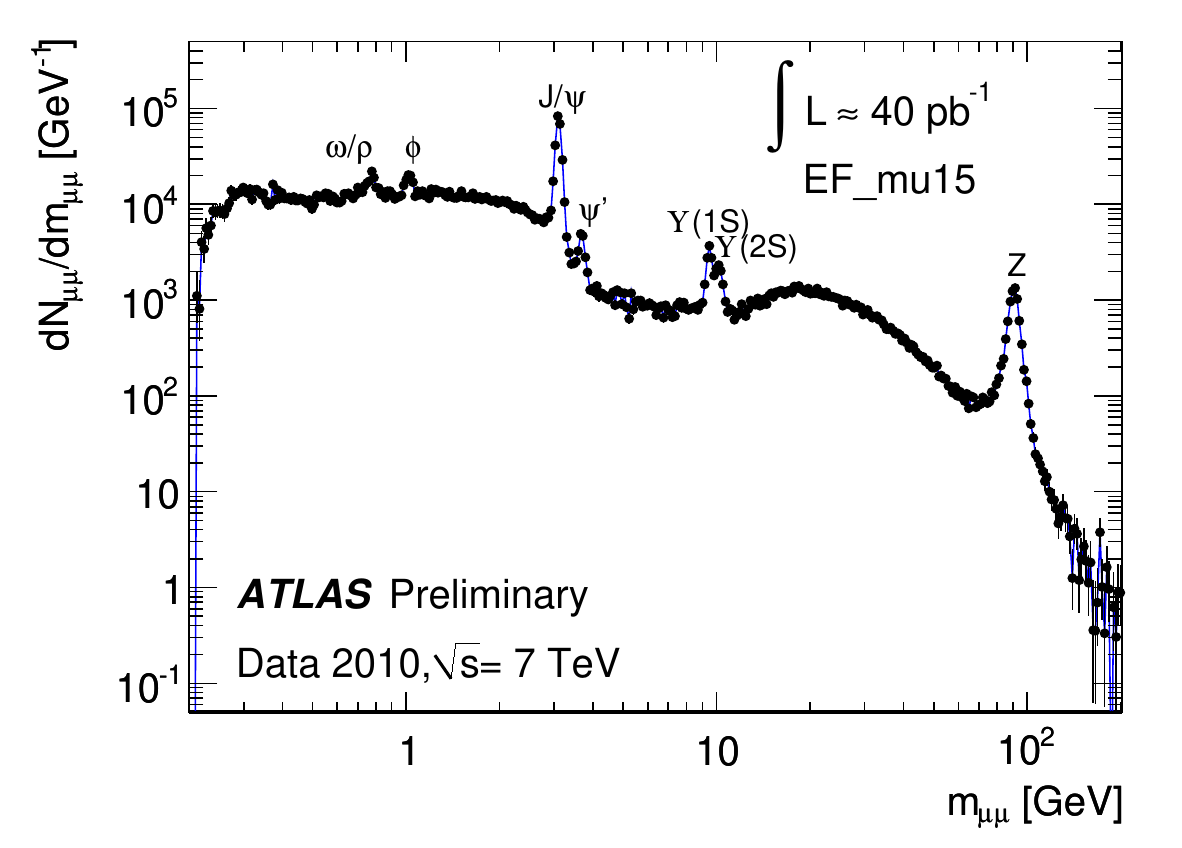}
\end{center}
\caption{\label{f:dimuons}
An example of dilepton spectrum in invariant mass, here  made exclusively from pairs of $\mu^+\mu^-$. The smooth background consists of random pairs and the peaks are due to pairs that come from decays of unstable particles, the names of the parent particles are indicated in the labels. Data from proton-proton collisions at centre-of-mass energy 7~TeV were collected by the ATLAS Collaboration in 2010. Figure copyright CERN for the benefit of the ATLAS Collaboration. CC-BY-4.0 license. \cite{ATLAS_public} 
}
\end{figure}
We illustrate a result of such a procedure in Fig.~\ref{f:dimuons}: this is a histogram of invariant masses of pairs $\mu^+\mu^-$, as obtained from data measured by the ATLAS Collaboration in  proton-proton collisions at centre-of-mass collision energy of 7~TeV. The smooth-curved shape of this invariant-mass spectrum is due to randomly paired muons. The labelled peaks are due to decays of various unstable particles, with their names listed in the labels. The observation or even discovery of an unstable particle is declared if the peak is clearly above the background and cannot result from a random statistical fluctuation of the background.

\begin{figure}[t]
\begin{center}
\includegraphics[width=0.42\textwidth]{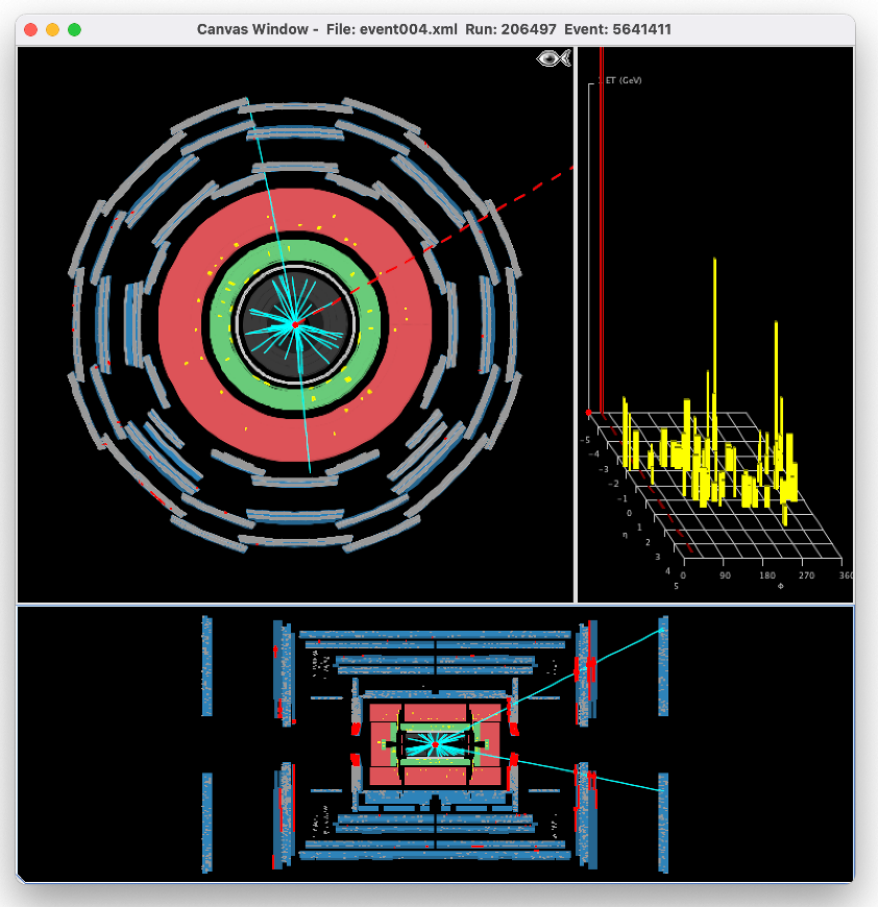}
\includegraphics[width=0.55\textwidth]{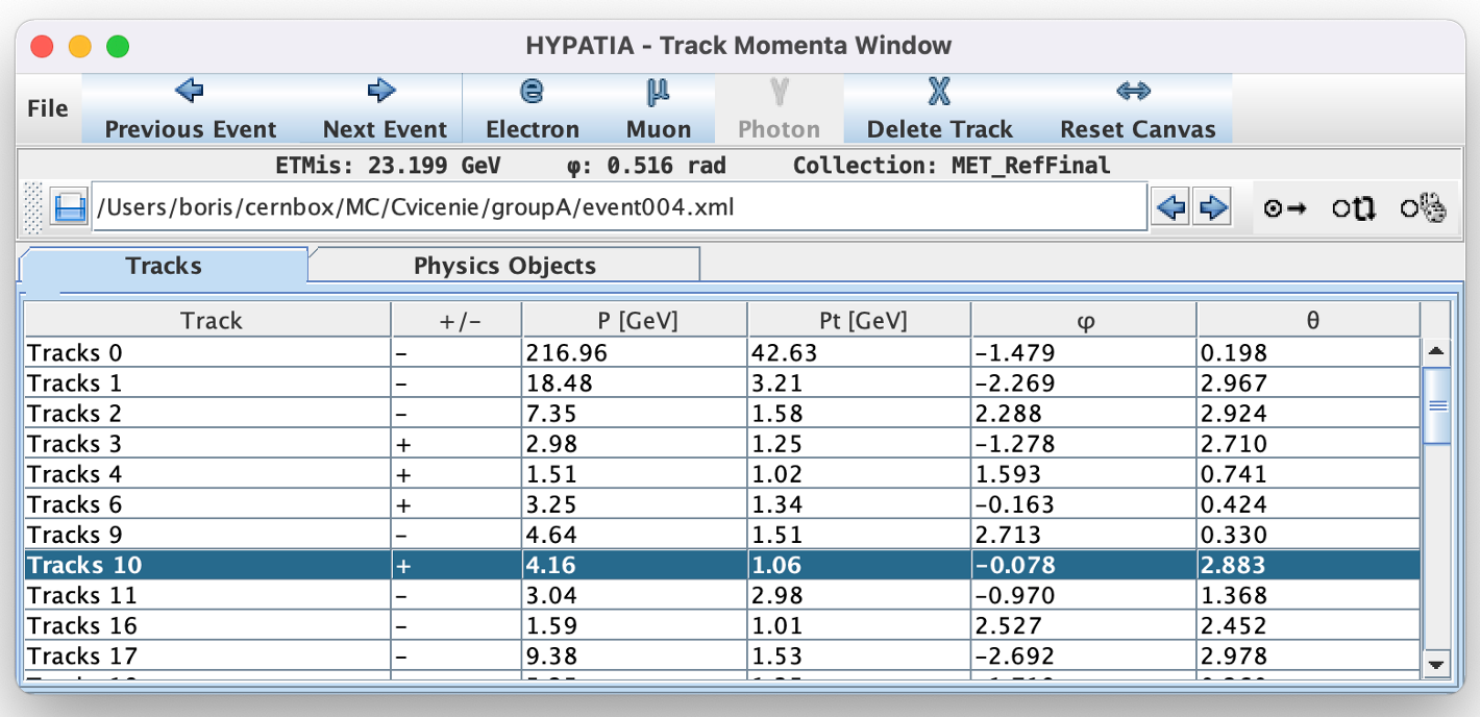}
\end{center}
\caption{\label{f:HYP}
The detector and analysis simulation software HYPATIA, used for the analysis of events. Left:  the canvas window which displays the top-view and the side-view of the detector, together with the histogram showing the angular distribution of energy deposition. The two tracks passing from the interaction point at the centre all the way to the outermost parts of the detector were generated by muons. Right: the track momenta window, which shows the numerical data for all tracks that are visualised in the canvas window.  }
\end{figure}

\begin{figure}[t]
\begin{center}
\includegraphics[width=0.325\textwidth]{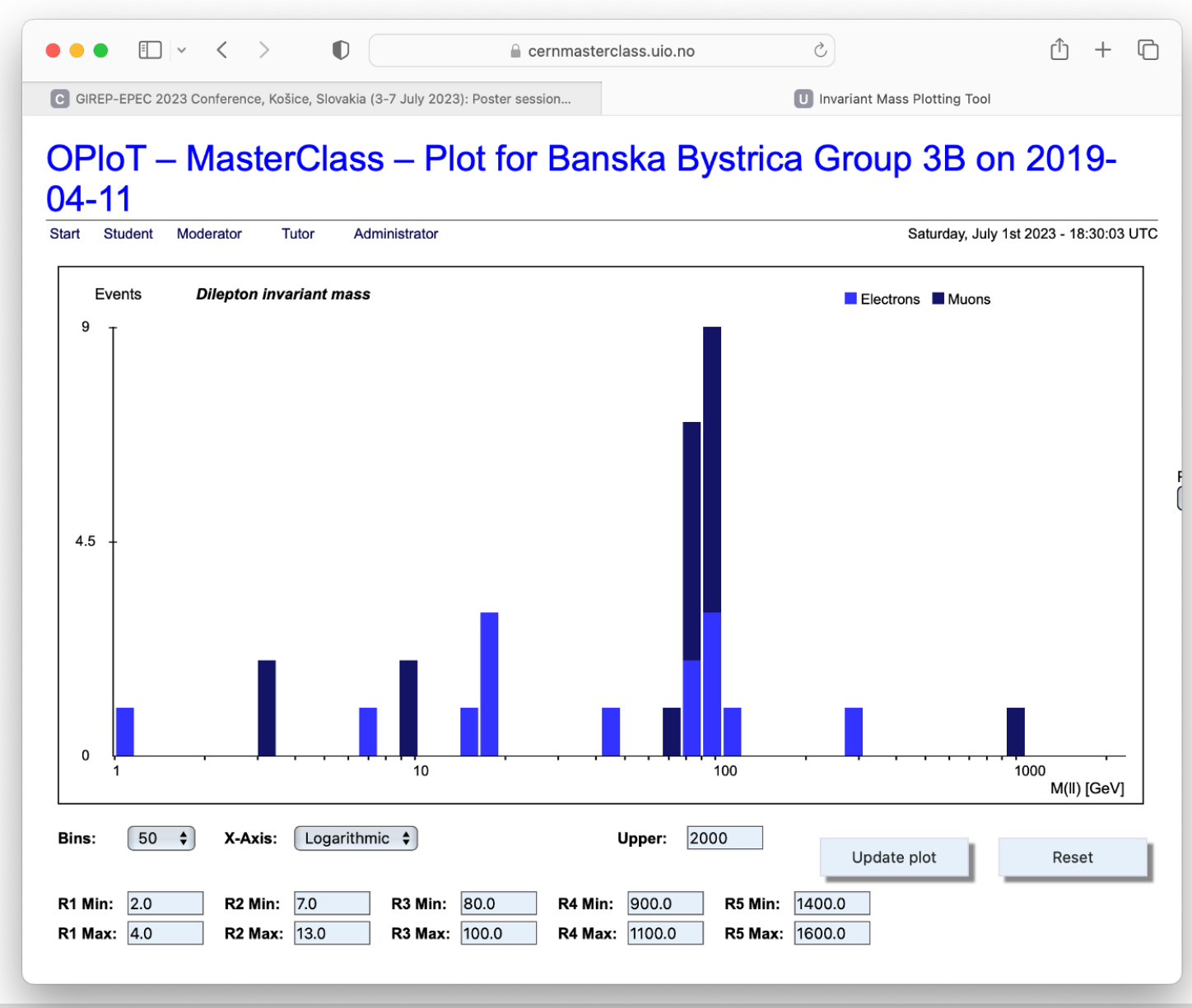}
\includegraphics[width=0.325\textwidth]{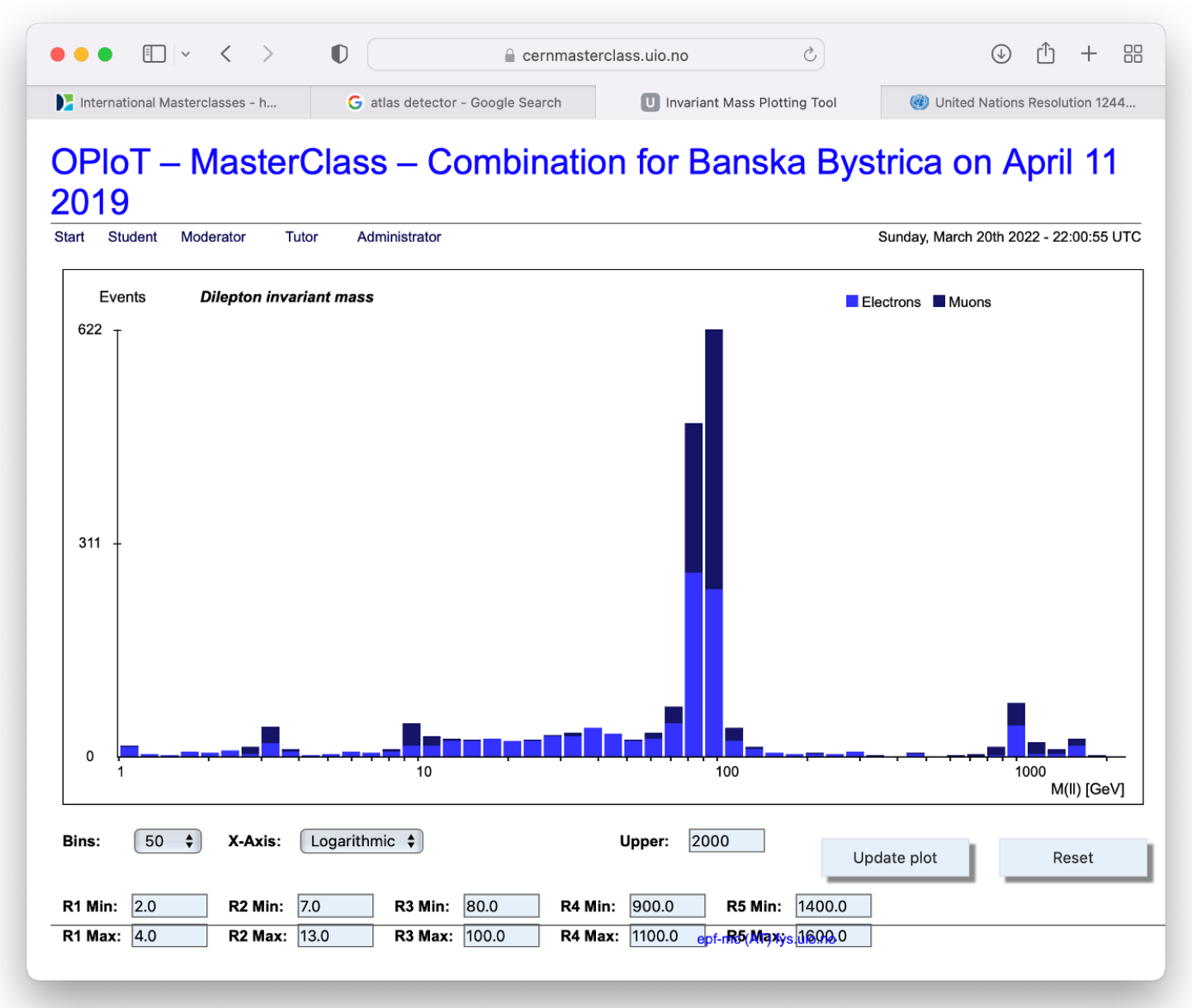}
\includegraphics[width=0.325\textwidth]{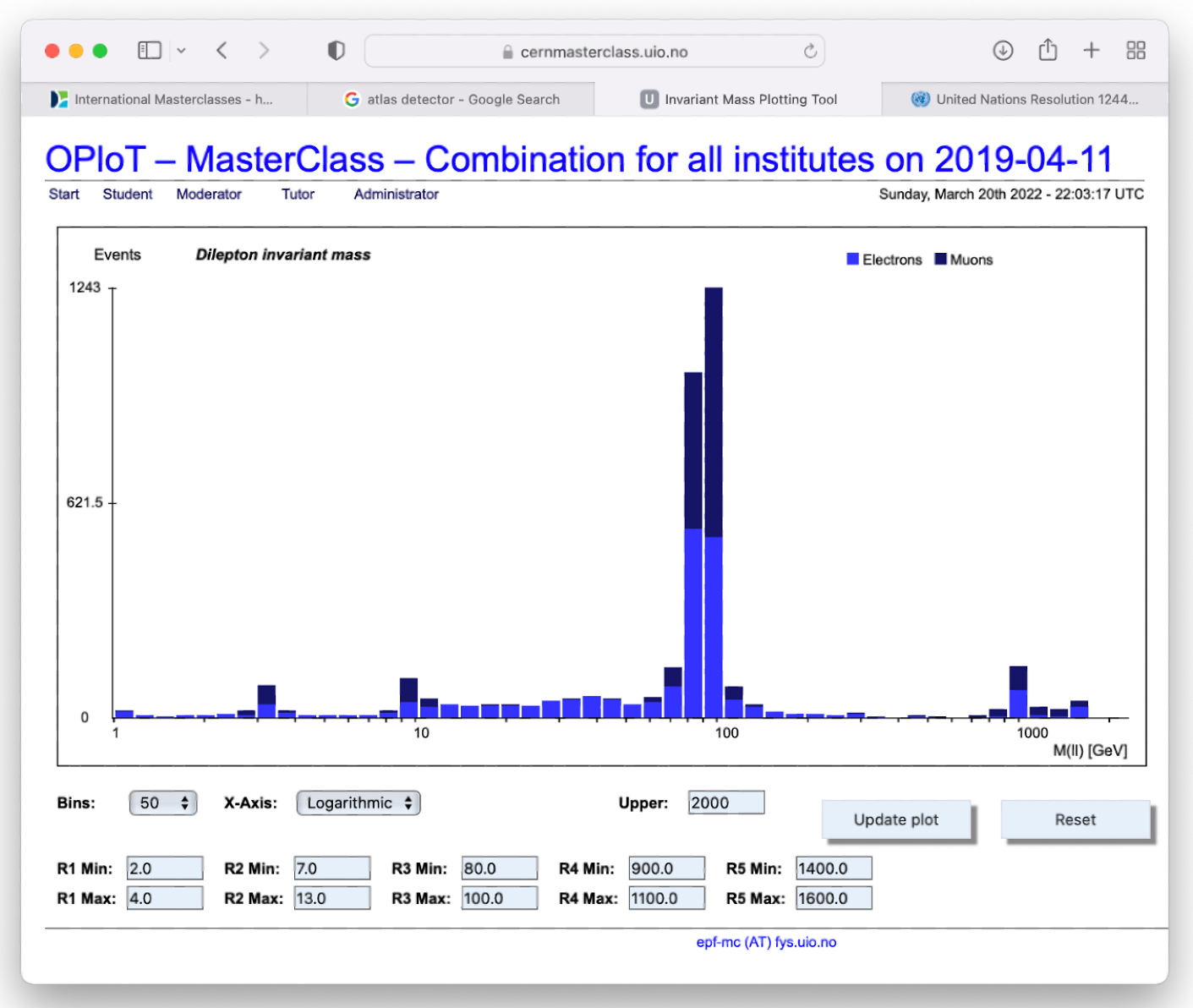}
\end{center}
\caption{\label{f:zhist}
Screenshots with the histograms of invariant masses obtained from analysing various numbers of collision events. Left: 50 events analysed by one participant; middle:  all events analysed by all participants of IMC in Bansk\'a Bystrica on 11.4.2019; right: all events analysed by all participants of IMC Europe-wide on 11.4.2019. The peak of the Z-boson is visible at 91~GeV. Other visible peaks are at 3.1~GeV ($J/\psi$ meson), 9.46~GeV ($\Upsilon$ meson), and 1000~GeV (an anticipated  $Z'$ particle,  added to the data).  }
\end{figure}

The task for the participants is to review 50 collision events and identify pairs of $e^+e^-$ or $\mu^+\mu^-$. 
For this, software tool HYPATIA is used, which visualises the traces of particles in the detector (Fig.~\ref{f:HYP}).
%
%
HYPATIA was originally developed as a tool for training doctoral students. It visualises signals of particles in the ATLAS detector. 
In IMC, a limited number of its functionalities is used. 
After finding the pairs one just needs to click on the shown tracks and the software calculates the invariant mass of the pair automatically.
After examining all 50 collision events, the extracted invariant masses are imported into another tool that produces histograms, so that the pupil obtains his/her histogram from the analysis.
An example  of such histogram is shown in Fig.~\ref{f:zhist}.
%
%
The tool also allows to merge results from all participants at one institution and later also results from all institutions participating in that day.
We show in Fig.~\ref{f:zhist} how the peak at the value of 91~GeV, which is the mass of the Z-boson, increases its clarity as more and more data are included in the analysis. 
This is an important lesson which the participants learn thanks to the collective effort coordinated in the exercise.

\section{Pandemic edition of Masterclasses}
\label{s:pand}

It is an important feature of an event like Masterclasses that the interested pupils are hosted directly at the research institutions. 
The venue contributes to the unique experience of working on a research task in high-energy physics and exposure to the work of physicists. 
Restrictions imposed on public gatherings due to pandemics then severely restrain some of the main ingredients of an IMC event. 
In such a unprecedented situation, there were no recommendations from IPPOG, which usually coordinates the local events. 
Most of the IMC events in Slovakia and the Czech Republic had to be cancelled in 2020 and 2021. 
(In fact, in one of the institutions we were lucky to host the event on the last day before closing of the schools, while in another the Masterclasses have been called off just the evening before the day.)
In order to uphold the tradition, we organised in Slovakia both years a nation-wide online event. 
There were several reasons to join the forces and set up one event together instead of organising IMC at each institution separately. 
An online event required new tasks to be accomplished and more manpower to prepare and run. 
On the other hand, while a traditional IMC event typically gathers a few tens of participants at each institution,  only about twenty of them connected to the online events. 

In 2020 we have used lectures recorded and posted on youtube from a regular IMC event in 2019\footnote{\href{https://youtu.be/LohhJBc-CeY?si=snH6AQEiPdpOqpCv}{https://youtu.be/LohhJBc-CeY?si=snH6AQEiPdpOqpCv}\\ \href{https://youtu.be/8x1s5coZIj0?si=IbJZvcNDdvlAQiDX}{https://youtu.be/8x1s5coZIj0?si=IbJZvcNDdvlAQiDX}}, as a replacement for the two morning lectures. 
In 2021, the two introductory lectures were delivered live online, and their recordings have been posted on youtube\footnote{\href{https://youtu.be/Nhf6iy3PRXU?si=i2vt80JRdKri1OcJ}{https://youtu.be/Nhf6iy3PRXU?si=i2vt80JRdKri1OcJ}\\
\href{https://youtu.be/nCuHaGfHnng?si=ofyp6UopTXc5kPDP}{https://youtu.be/nCuHaGfHnng?si=ofyp6UopTXc5kPDP}}. 
For these postings we have used the youtube channel \emph{Svet \v{c}ast\'ic} (The World of Particles). 
It is a part of the portal devoted to particle physics outreach in Slovakia. 

The exercises were done by the participants at their homes. 
The first challenge was the installation of the software along with the necessary data sets. 
Even though it is based on java and runs usually easily, there always appear individual cases of problems. 
To help the participants to solve them,  we established online help with graduate students volunteering to be available for a help with troubleshooting. 
The instructions\footnote{\href{https://youtu.be/gQXVjCdoc7A?si=3d3aJI35ugimtC6C}{https://youtu.be/gQXVjCdoc7A?si=3d3aJI35ugimtC6C}} and commented data analysis\footnote{\href{https://youtu.be/nEbTBEmg4Gg?si=swnKt5zBMNaFw-NB}{https://youtu.be/nEbTBEmg4Gg?si=swnKt5zBMNaFw-NB}} of some data were screen-recorded and also posted on youtube.
Finally, during the designated time slot, consultants were available in videoconference rooms and ready to answer questions and help with problems at data analysis. 
The whole event was closed by common session for all participants in which the data analysis was summarised. 

The setup with participants connecting individually has been also allowed by IPPOG later for the videoconferences on international level.

\section{CASCADE projects}
\label{s:cascade}

The immediate impact of IMC is usually large: participants become excited and motivated to learn even more about particle physics. 
It is appropriate then to foster the interest. To this end, mainly in Slovakia the CASCADE competition is being organised. 

The scheme is to have the interested and motivated pupils to prepare educational content on (particle) physics topics and present it to their peers. 
In this way, through the participating pupils we can reach a few hundred teenagers which we could hardly manage on our own. 

Before the pandemic, the task was to prepare and present a talk in front of fellow students. 
It was a competition of three-members teams. 
The talks were recorded and the records were sent to a jury who either selected the winners or invited the best teams to participate in the country-wide finals. 
The evaluation was based on both the content and the quality of the presentation itself. 

This was no longer doable after 2020. 
At that point, CASCADE was made a competition for individuals and the tasks were to prepare a video that can be posted online. 
The motivation was---in addition to the impossibility of in-person meetings---that online videos generally may have higher impact. 
Again, the jury evaluated the actual content as well as the quality and attractiveness of the videos.

\section{The impact of Masterclasses}
\label{s:impact}

An initiative like Masterclasses that has been running very successfully since almost 20 years and hosted tens of thousands of pupils worldwide should have influenced a generation of young people. 
One would expect that there could be young researchers who have participated in IMC and were influenced by it. 
Another question could be, if and how IMC have influenced those who chose careers outside of particle physics or even science. 

To answer these questions, two anonymous surveys were performed by means of google questionnaires in spring 2023. In \emph{Survey 1} we profited from paper questionnaires that were collected between 2011 and 2015 immediately after the IMC events at all participating Slovak institutions, and once in the Czech Republic. 
At that time, we were interested in the immediate effect of Masterclassees, we analysed the quality of their individual parts, and asked about the motivation of participants to partake. 
Finally, an optional request was made to enter the email address, so that we can approach the respondents later. 
Now, these email addresses were used to invite former IMC participants to respond to our survey. 
Out of about 1000 collected paper questionnaires, in the end 71 responses were received. 
Generally, the survey focused on how the careers of the participants evolved  after the experience with IMC and how they changed their attitude concerning science and especially particle physics. 

In \emph{Survey 2} we solicited answers from young high-energy physicists in the Czech Republic and Slovakia. 
In anticipation that some share of them has participated in IMC it asked about how the experience influenced them in the choice to become physicists. 
Out the potential community of about 125 people 18 responses have been collected. 

The results have been summarised in a dedicated publication \cite{Kekelakova:2023sjl}. 
Here, we highlight some of the results that have been collected there.

%
\begin{figure}[t]
\begin{center}
\includegraphics[width=0.34\textwidth]{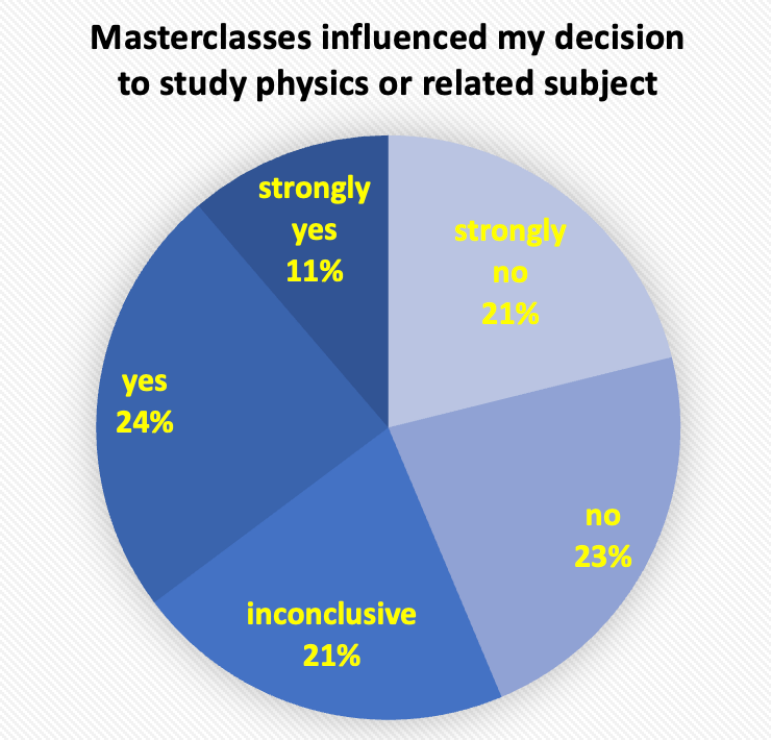}
\includegraphics[width=0.305\textwidth]{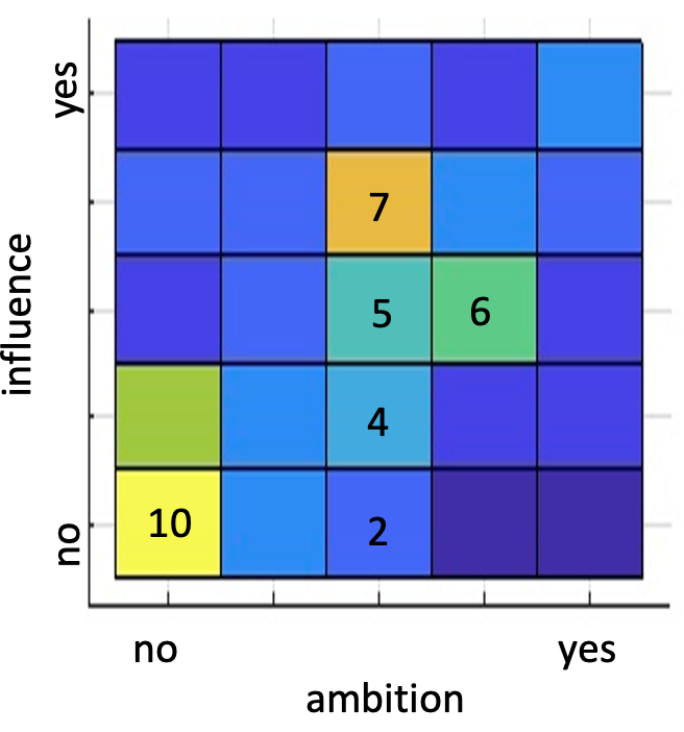}
\includegraphics[width=0.34\textwidth]{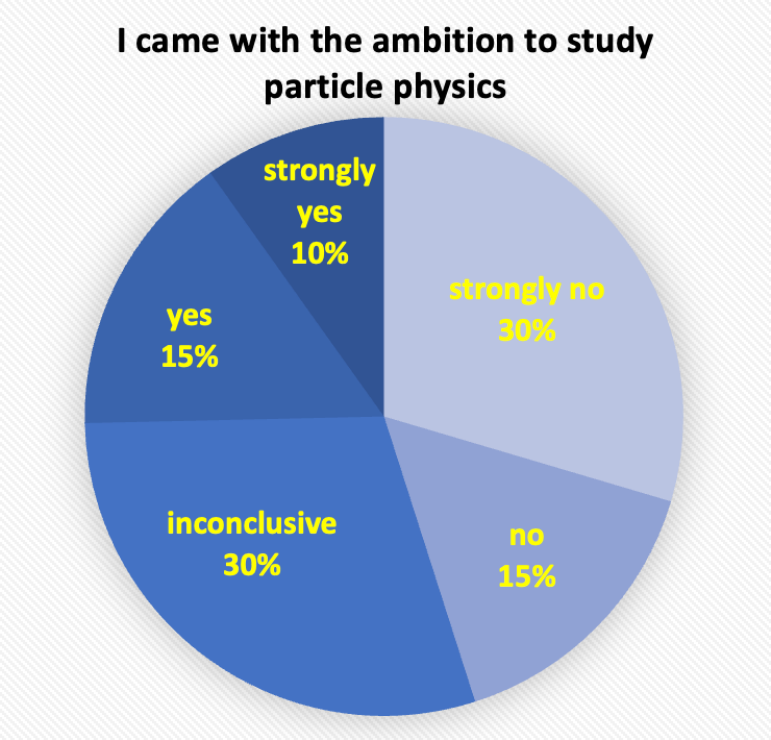}
\end{center}
\caption{
Left: Responses to the question: ``Masterclasses have influenced my decision to study physics or related subject.'', on 5-point Likert scale. Right: Responses to the question ``I came to Masterclasses with the ambition to study particle or nuclear physics.'', on 5-point Likert scale.  Middle: 3-dimensional histogram with the answers to both previous questions. 
\label{f:aicorrel}
 }
\end{figure}
%
In Fig.~\ref{f:aicorrel} we show that about a quarter of participants came to IMC with the ambition to later study particle physics. 
On the other hand, 45\% of them had some other plans. 
Nevertheless, 35\% of the participants claim that IMC  influenced them to later choose physics or related subject. 
To gain more differential insight into this issue, we also show how the answers to these two question are correlated. 
There is a most clear correlation: those participants who clearly had no ambition to later study particle physics were not influenced in their decision. 
However,  an interesting nudge effect is there for those who were not strictly against such an idea. 
Some people that were ambivalent about choosing particle physics report positive influence to later go into particle physics research. This is good news for the IMC organisers: while you do not convince people not planning to do physics later (frankly---it would be naive to expect this), you may support the motivation for those who would not decide so without the IMC experience. 

%
\begin{figure}[t]
\begin{center}
\includegraphics[width=0.28\textwidth]{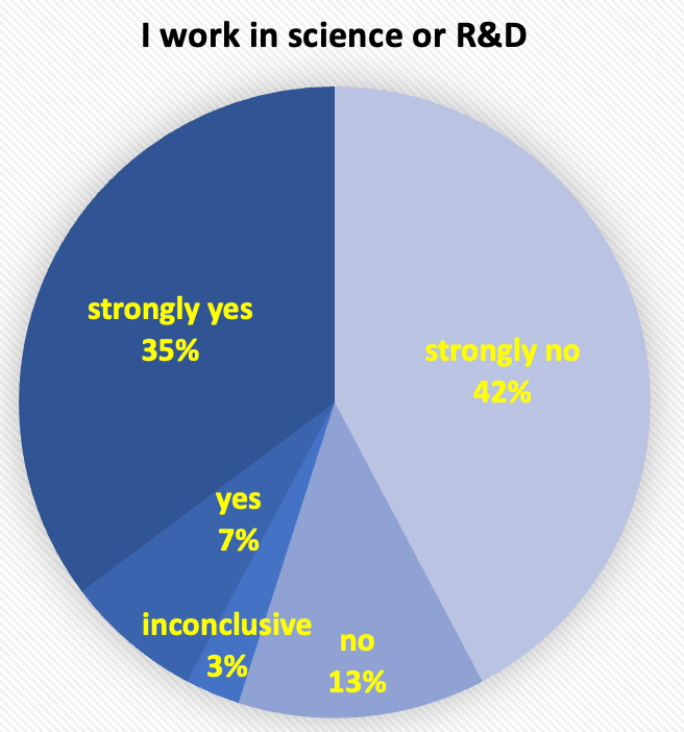}
\includegraphics[width=0.7\textwidth]{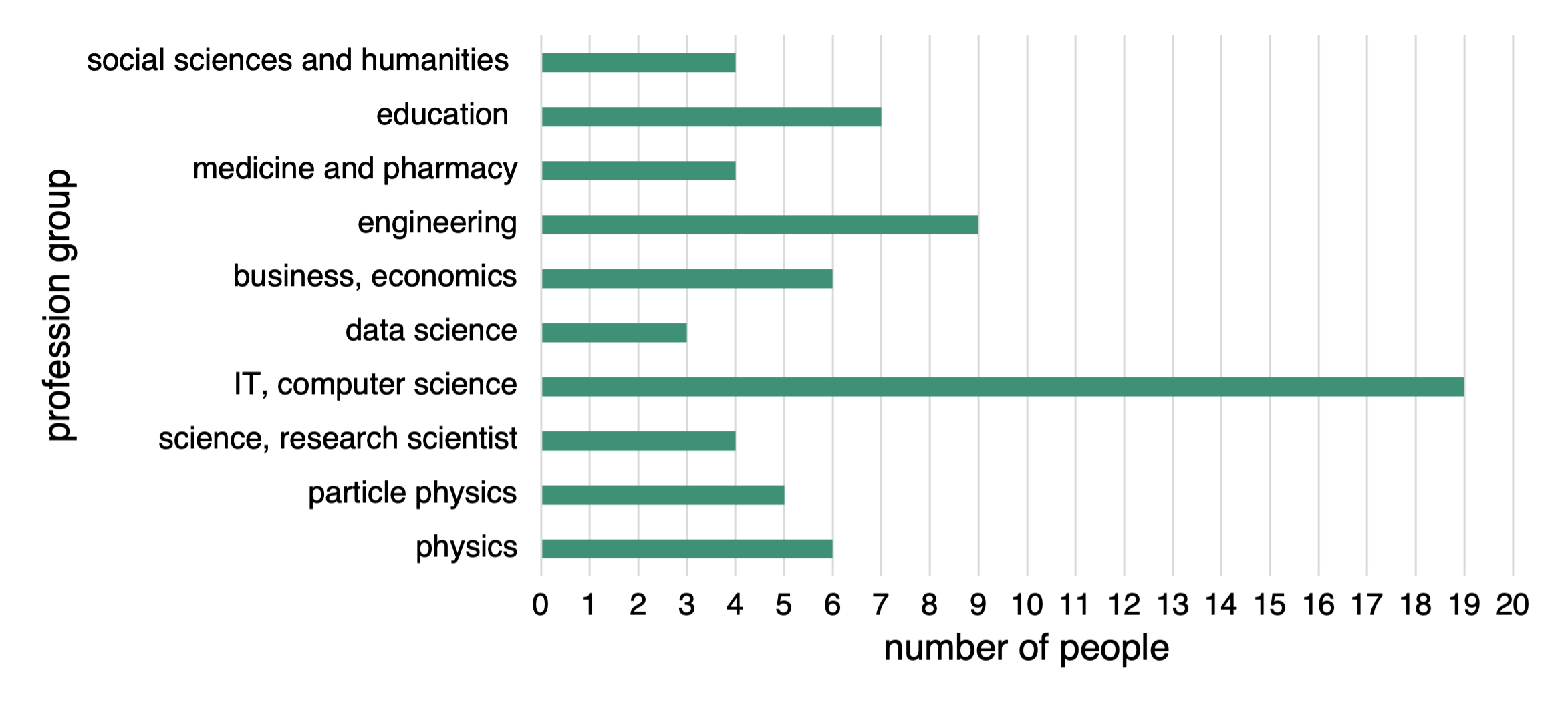}
\end{center}
\caption{Left: The distribution of respondents to Survey 1 to professions in Science and/or Research\&Development and those outside this domain. Right: Histogram of different professional groups of respondents to Survey 1. 
\label{f:prof}
 }
\end{figure}
%
What are the careers that the respondents chose? This is shown in Fig.~\ref{f:prof}. 
In fact, more than a half of the respondents have jobs outside science or research and development (R\&D). 
About 42\% of the participants identify their jobs like belonging to this domain. 
A detailed analysis of the professions reveals that by far most of the respondents finally build up their career in the IT branch. 
The other most populated groups of professionals---with a large gap---are  engineers and education professionals, which are presumably mostly STEM teachers. 
Physics is populated as much as business and economy, and they are closely followed by particle physicists. 
We conclude, that the ideas conveyed in the IMC events become distributed over a wider spectrum of different professions within the society.

Since some of the current respondents opted to leave also their email addresses, we can compare the responses in 2023 with those right after the IMC event. In total, there were 36 responses which are available for this kind of analysis. It is interesting to compare the detailed plans about the profession (``What would you like to do in the future?'') with the actual reality (``The content of my current employment or study is:''). 
%
\begin{figure}[t]
\begin{center}
\includegraphics[width=0.5\textwidth]{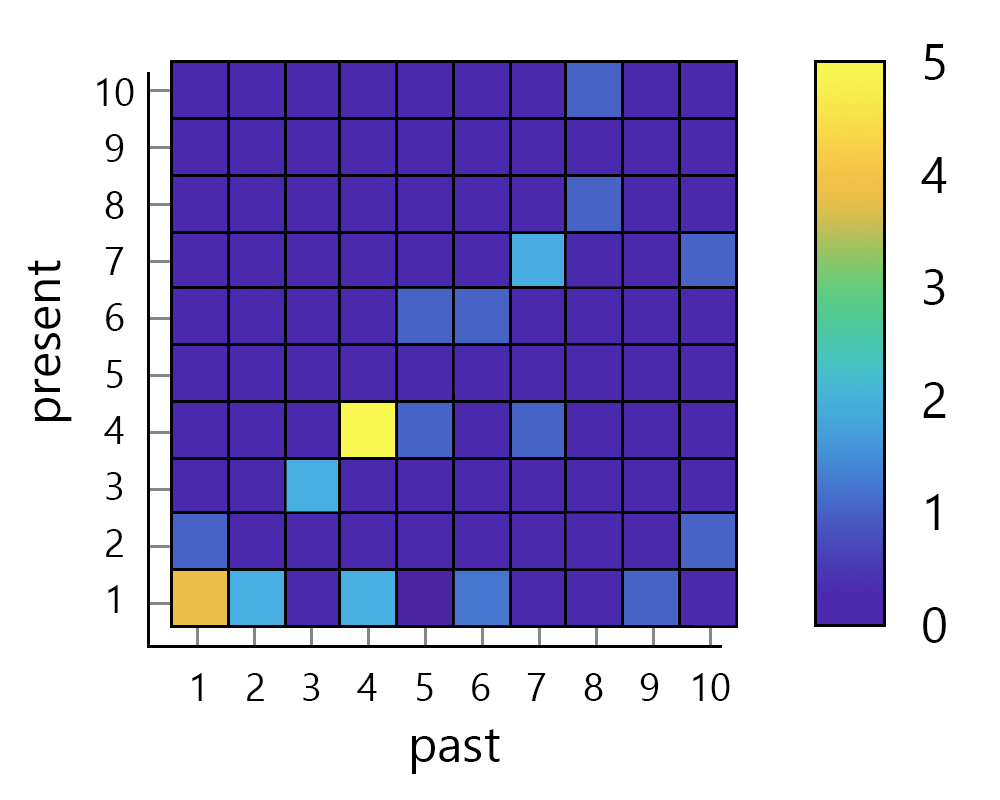}
\end{center}
\caption{The correlation between the profession planned just after the respondents have participated in an IMC event (denoted `past') and the actual present career (denoted `present'). The coding of the profession groups: 1 physics; 2 particle physics; 3 science, research scientist; 4 IT, computer science; 5 data science; 6 business, economics; 7 engineering; 8 medicine and pharmacy; 9 education; 10 social sciences and humanities. 
\label{f:beforeandafter}
 }
\end{figure}
%
The comparison is presented in Fig.~\ref{f:beforeandafter}. The professions were grouped into 10 groups which are described in the Figure caption. 
If there was no change in plans, all entries would be arranged along the diagonal. 
Indeed, such a pattern seems to be visible. 
The brightest peak corresponds to IT professionals, and the second brightest to physicists. 
Some other features can be observed, as well, though they should be taken with a grain of salt as these conclusions are based on very limited statistics. 
The group of present physicists collects also a few people who had different plans, originally. 
Similar, though weaker collection is visible for the IT-professionals.

%
\begin{figure}[t]
\begin{center}
\includegraphics[width=0.45\textwidth]{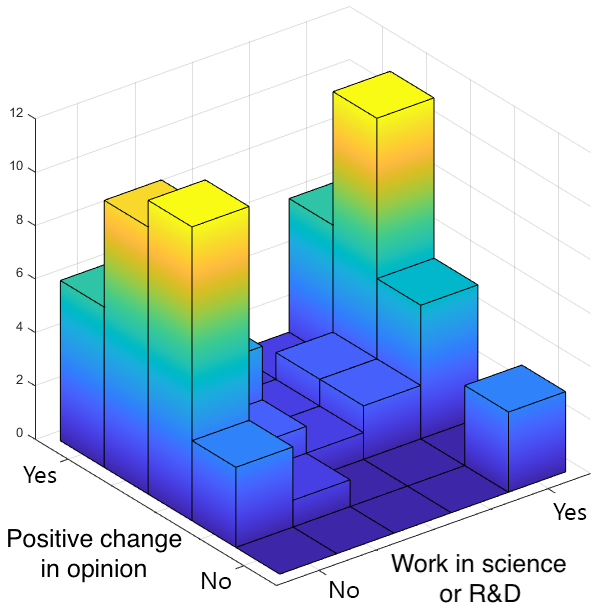}
\end{center}
\caption{Three dimensional histogram of answers to the questions: ``Thanks to Masterclasses I positively changed my opinion about science, research, and physics.'' (5-point Likert scale) and ``I contribute to the development of science---I am scientific associate, or I work for a company focussed on R\&D, or I want to work in this field after I finish my study.'' (5-point Likert scale). 
\label{f:q5q7}
 }
\end{figure}
%
Finally, we present in Fig.~\ref{f:q5q7} an analysis of the answers that reveal whether the respondents consider IMC to influence them positively in their opinion about science. 
The respondents  are basically divided into two groups: those who work in science or R\&D and those who do not. 
The scientists report a clear positive shift in the opinion about science. 
For the other group, the most frequent answer is ambivalent, but the distribution of answers is also tilted to the positive side. 

In summary, the survey indicates that IMC may have a positive nudge effect: they are supportive in decisions for science-oriented careers. 
Moreover, even for those who do not choose science careers, they help to spread the positive attitude towards science, and particle physics in particular.

\section{Conclusions}
\label{conc}

The large effort that has been invested in developing and organising outreach in form of non-formal education activities has proven to pay off after years of running it. 
The crucial aspects are analysis of real data and true international collaboration even during the IMC event. 

There is a group of scientists who have been influenced by IMC in their career decision phase and there is an impact on wider society concerning its attitude towards science. 

The exercises developed for IMC even found their ways, e.g., to freshmen labs at college levels, or pre-service teacher training. 

We would be happy if the  activity is inspirational also to other branches of physics and science in general. There are examples of Masterclasses in astronomy. 
The rich practical experience acquired meanwhile in particle physics, part of which we tried to document here, may be helpful in developing new masterclasses in other fields. 


\subsection*{Acknowledgements}

We thank Pavol Barto\v{s} for pointing us to Fig.~\ref{f:dimuons}.


\end{document}